# Monolithic GaN Systems Combining Non-Volatile Memory and Analog Computing via Area-Ratio-Engineered Ferroelectric AlScN Gate Stacks

*Hyeong Jun Joo[1], Jong Min Park[1], Kieran Barrett-Snyder[2], Brendan Hanrahan[3,*], and Geonwook Yoo[1, 4, *]*

**Corresponding authors. Email: brendan.m.hanrahan.civ@army.mil, gwyoo@ssu.ac.kr*

[1] Department of Intelligent Semiconductor, Soongsil University, Seoul 06938, Korea

[2] Fibertek Inc, Herndon, VA, USA

[3] DEVCOM Army Research Laboratory, Adelphi, MD, USA

[4] School of Electronic Engineering, Soongsil University, Seoul 06938, Korea

**Abstract**

Gallium nitride (GaN) transistors have become the platform of choice for power electronics and radio-frequency power amplifiers. To unlock capabilities beyond those of conventional GaN, integrating ferroelectric heterostructure has been considered toward memory, logic and reconfigurable systems. Here, we demonstrate ferroelectric GaN transistors employing an AlScN-based gate stack in which systematic engineering of the area-ratio ($S_{MIS}/S_{MFM}$) provides unified control over both memory and analog functionality. Precise modulation of the intermediate electrode length yields a record memory window of 27 V and a forward subthreshold swing of 27 mV/dec, driven by ferroelectric polarization reversal of a robust downward-polarization state pre-induced by two-dimensional electron gas. Low area-ratio devices ($S_{MIS}/S_{MFM}$ = 1, 2) achieve 4-bit multi-level cell operation and excellent spatial uniformity across a fabricated 4×4 array, benchmarking favorably against established silicon and oxide ferroelectric architectures. High area-ratio devices ($S_{MIS}/S_{MFM}$ = 4, 8) harness continuously tunable conductance states to demonstrate multi-state inverters and the first GaN-based ferroelectric frequency-to-voltage converter, delivering a linear frequency-voltage response across 0.5–500 Hz range with a conversion gain of 1.1 mV/Hz. This work establishes routes towards monolithically integrated GaN systems that combine non-volatile memory and analog signal processing on a platform inherently suited to high-power and radio-frequency applications.

Gallium nitride (GaN) is a key wide-bandgap semiconductor that has been extensively investigated for high-power switching[1-3] and radio-frequency amplification[4-6]. As discrete components, GaN transistors have reached commercial maturity. However, every deployed GaN power or RF module still requires silicon-based peripheral circuits for non-volatile configuration storage, gate driving, and control logic[7,8]. This reliance on silicon constrains the path towards monolithic GaN integrated systems. As a radical approach, integrating ferroelectric structure directly into the GaN gate stack not only offers a route to overcome this bottleneck by enabling on-chip non-volatile memory but also imparts high functionality to foster GaN-based emerging devices.

Early efforts exploited various ferroelectric materials, including van der Waals 2D materials, perovskites, and hafnium-based fluorites to reconfigure GaN HEMT threshold voltages. These approaches yielded proof-of-concept demonstrations in reconfigurable RF circuits[9-11], multi-state logic[12-14], and amplifier-less actuation[15]. However, exfoliated 2D ferroelectric layers introduced process variability incompatible with wafer-scale integration; PZT and hafnium-based devices achieved relatively narrow memory windows for multi-bit analog storage. Moreover, the low coercive fields can make the programmed states susceptible to unintentional switching under the internal field generated by high-density two-dimensional electron gas (2DEG) as well as other spontaneous polarization within the epitaxial layers.

Recent transition to reactive sputtered wurtzite AlScN resolved both the material compatibility and polarization magnitude constraints[16-20]; Recessed gate structure enabled enhancement-mode operation and improved reliability by suppressing the depolarization field imposed by the residual AlGaN barrier[21,22]. Despite these advances, unique material properties of AlScN such as large remnant polarization exceeding 100 μC/cm$^2$ and coercive field of ~ 5 MV/cm have not been fully exploited, and thus fundamental challenges remain: Simultaneously maximizing the memory window, achieving sub-thermal switching, and enabling multi-state analog programmability within a single scalable gate stack structure. Toward this, voltage drop across the AlScN layer needs to be precisely controlled and amplified, and conventional ferroelectric gate stacks offer no structural degree of freedom to optimize the distribution.

In this work, we demonstrate that systemically scaling the area-ratio ($S_{MIS}/S_{MFM}$) of a metal-ferroelectric-metal-insulator semiconductor (MFMIS) gate stack controls capacitive voltage partition between the ferroelectric AlScN and insulator layer of ferroelectric GaN HEMTs. The design strategy achieves a record memory window of 27 V, a forward subthreshold swing of 27 mV/dec driven by 2DEG-pinned polarization reversal, and robust 4-bit multi-level cell operation with a 4×4 memory array. Simultaneously, we implement reconfigurable multi-state inverters and the first GaN-based ferroelectric frequency-to-voltage converter (FVC) within a single scalable platform. These results establish area-ratio engineering as a practical design principle for FeHEMTs with AlScN gate stack, bridging high-density non-volatile memory and precision analog signal processing on a single GaN platform.

Figure 1a depicts the cross-sectional schematic of fabricated FeHEMT with the MFMIS gate stack. Figure 1b shows an optical microscope image of a representative, fabricated FeHEMT with an area-ratio of 2. As shown in Figure S1a, tuning the length ($L_{Pt} = L_{MIS}$) of intermediate Pt electrode layer at the fixed top gate length ($L_G = L_{MFM}$) allows tuning the $S_{MIS}/S_{MFM}$ ratio (see Figure S1b for an area-ratio of 8), and thus directly modulates the electrostatic coupling between the MFM and MIS capacitors of the gate stack. The area-ratio ($S_{MIS}/S_{MFM}$) is a design parameter for engineering the internal voltage partitioning to efficiently enhance the potential drop across the ferroelectric MFM layers. The structure and elements of the gate stack are confirmed through cross-sectional transmission electron microscopy and energy-dispersive X-ray spectroscopy maps (Figure 1c). Moreover, utilization of the Pt electrode can serve as a structural template that promotes highly c-axis oriented crystalline growth of the AlScN layer, which is imperative for robust ferroelectricity[23,24]. Under electrostatic boundary conditions of a series capacitor network, applied gate bias ($V_{GS}$) is distributed according to the voltage division, $V_{GS} = V_{MFM} + V_{MIS}$. The potential drop across the ferroelectric MFM capacitor is determined by:

$$V_{MFM} = V_{GS} \times \frac{1}{1 + C_{MFM}/C_{MIS}}$$

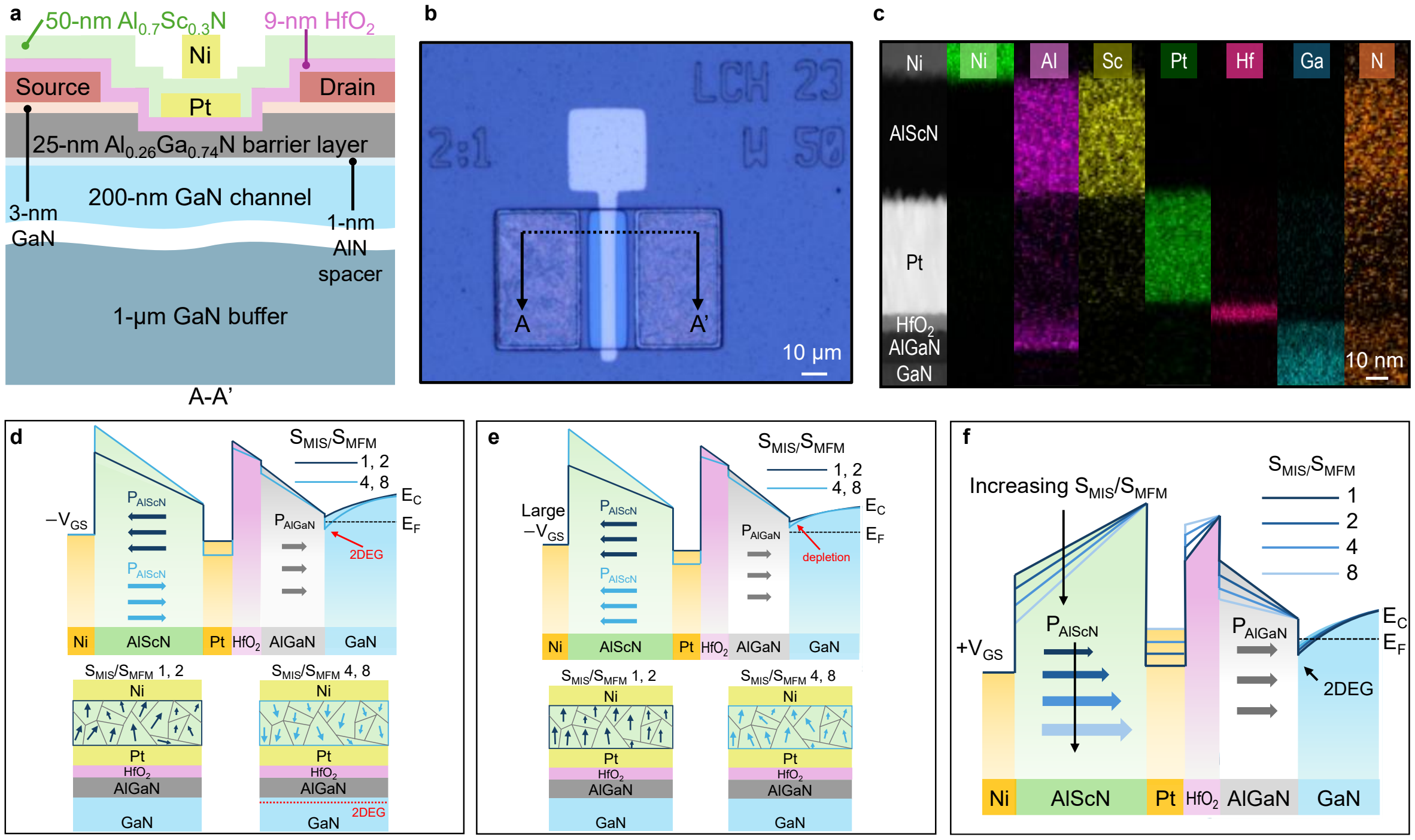

**Figure 1. Structural characterization and energy band analysis of FeHEMTs. (a)** Cross-sectional schematic of the proposed MFMIS FeHEMT. **(b)** Optical microscope image of the fabricated FeHEMT with an area-ratio of 2. **(c)** Cross-sectional transmission electron microscopy (TEM) image and corresponding energy-dispersive X-ray spectroscopy (EDS) mapping of the MFMIS gate stack. **(d), (e)** Schematic of energy band diagrams and corresponding polarization domains of the FeHEMTs for low (1 and 2) and high (4 and 8) area-ratios ($S_{MIS}/S_{MFM}$) under (d) negative $V_{GS}$ and (e) large negative $V_{GS}$ in the off-state. **(f)** Energy band diagrams of the FeHEMTs for all area-ratios in the on-state (positive $V_{GS}$).

And capacitance ratio is formulated as:

$$\frac{C_{MIS}}{C_{MFM}} = \frac{S_{MIS}}{S_{MFM}} \times \frac{d_{MFM}}{d_{MIS}} \times \frac{\varepsilon_{MIS}}{\varepsilon_{MFM}}$$

where $\varepsilon_0$, $\varepsilon_r$, S, and d denote vacuum permittivity, relative dielectric constant, area, and thickness, respectively. Based on C-V measurement, relative permittivity ($\varepsilon_r$) of $HfO_2$ and AlScN was extracted to be 19.3 and 24, respectively. Considering these values, an area-ratio of $S_{MIS}/S_{MFM} = 8$ substantially yields $V_{MFM} \approx 0.97\ V_{GS}$. This signifies near-ideal voltage transfer efficiency, where the applied gate potential is predominantly dropped across the AlScN.

In this regard, energy band diagrams and corresponding polarization domains differ depending on the area-ratios. For low area-ratios (1 and 2), a moderate negative $V_{GS}$ is sufficient to induce electrostatic pinch-off and deplete the 2DEG, consistent with standard FeHEMT operation (Figure 1d). In contrast, high area-ratio devices (4 and 8) exhibit incomplete depletion under the same bias and require substantially larger negative $V_{GS}$ to achieve full 2DEG depletion and polarization charge ($P_{AlScN}$) switching (Figure 1e). This behavior is attributed to two coupled mechanisms in the proposed FeHEMTs: First, the inherently high sheet density of the 2DEG at the AlGaN/GaN heterojunction generates a strong built-in electric field that pre-aligns and pins the AlScN polarization downward. Second, the longer $L_{Pt}$ increases total 2DEG charge ($Q_{2DEG}$) that must be depleted for channel-off, requiring a substantially larger negative bias to overcome both increased $Q_{2DEG}$ and pinned polarization. For positive $V_{GS}$, downward band bending of the AlScN layer becomes progressively steeper as the area-ratio increases under same $V_{GS}$ due to increased potential drop across the AlScN layer (Figure 1f). This enhanced effective gate bias across the AlScN layer directly correlates with $P_{AlScN}$. P-V hysteresis of the standalone MFM capacitor (Figure S1c) shows that higher applied fields drive minor sub-loops toward full saturation. Consequently, $P_{AlScN}$ is amplified for the high area-ratio devices at the same $V_{GS}$.

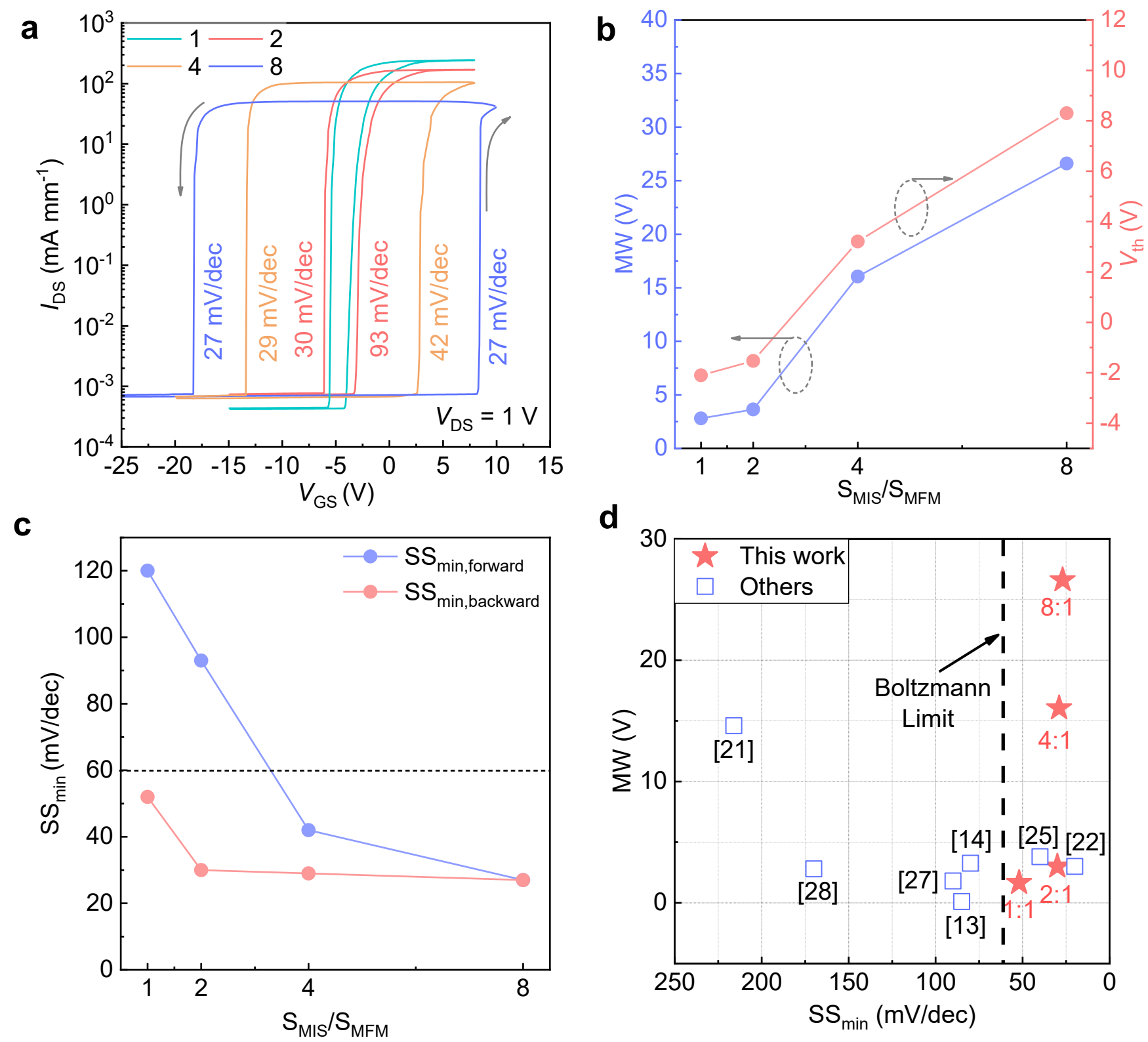

**Figure 2. Electrical characterizations and performance benchmarking of FeHEMTs. (a)** Bidirectional transfer sweeps for area-ratios from 1 to 8. **(b-c)** Extracted memory window (MW), threshold voltage ($V_{th}$), and forward and backward minimum subthreshold swing ($SS_{min}$) as a function of area-ratio. **(d)** Benchmarking of MW and $SS_{min}$ compared to previously reported GaN-based FeHEMTs.

Figure 2a shows bidirectional transfer curves of the FeHEMTs for the area-ratio from 1 to 8. Extracted key parameters are summarized in Figure S2a and Figures 2b and 2c. Increasing the area-ratio progressively degrades both the on-current and the on/off current ratio (Figure S2a). The geometric extension of gate-controlled channel length ($L_{Pt}$) introduces additional series resistance within the 2DEG channel. Figure 2b reveals a pronounced positive shift in the threshold voltage coupled with increased memory window (MW) with higher area-ratios. Under negative gate bias, the ferroelectric polarization aligns upward. According to the capacitive voltage division, an increased area-ratio enhances the effective electric field across the AlScN layer, thereby facilitating full polarization switching rather than minor-loops. Consequently, a higher gate bias is required to overcome the negative polarization charges and accumulate electrons in the depleted 2DEG channel, which inherently translates to a positive threshold voltage ($V_{th}$) shift and an increased MW. Figure 2c presents minimum subthreshold swing ($SS_{min}$) extracted from both forward and backward sweeps. As the area-ratio increases, $SS_{min}$ decreases in both sweep directions. For low ratio devices (1 and 2), sub-60 mV/dec is observed only during the backward sweep, which is consistent with typical behaviors of FeHEMTs[9,25]. However, high ratios devices (4 and 8) also exhibit sub-60 mV/dec during the forward sweep that has not been previously reported in GaN-based ferroelectric transistors.

To elucidate this anomaly, transfer characteristics were systematically analyzed by modulating the initial sweep voltage ($V_{start}$) in Figures S2b-S2e. For the low ratios (1 and 2), the electrostatic pinch-off to deplete the 2DEG is achieved around $V_{th}$ (Figures S2b and S2c). In contrast, the high ratio devices (4 and 8) require a larger negative bias to deplete the 2DEG channel (Figures S2d and S2e). Thus, accumulated $Q_{2DEG}$ exerts a significant electrostatic screening effect. This screening induces a strong built-in electric field that pre-aligns and pins the AlScN polarization toward the channel. When sweeping around $V_{th}$, therefore, the 2DEG channel is not depleted, necessitating a significantly larger negative bias. This behavior indicates that the AlScN polarization is already strongly aligned downward before the sweep begins. Consequently, during the forward sweep, dynamic reversal of the strong downward polarization triggers a transient negative capacitance effect[26]. This provides an internal voltage amplification delivering sub-thermal switching of $SS < 60$ mV/dec even in the forward sweep. Figure 2d presents benchmarking of key parameters against state-of-the-art FeHEMTs. The proposed AlScN-based MFMIS architecture not only exhibits competitive metrics at low ratios but also successfully overcomes trade-off by demonstrating unprecedented superiority in both MW and $SS_{min}$ at scaled area-ratios.

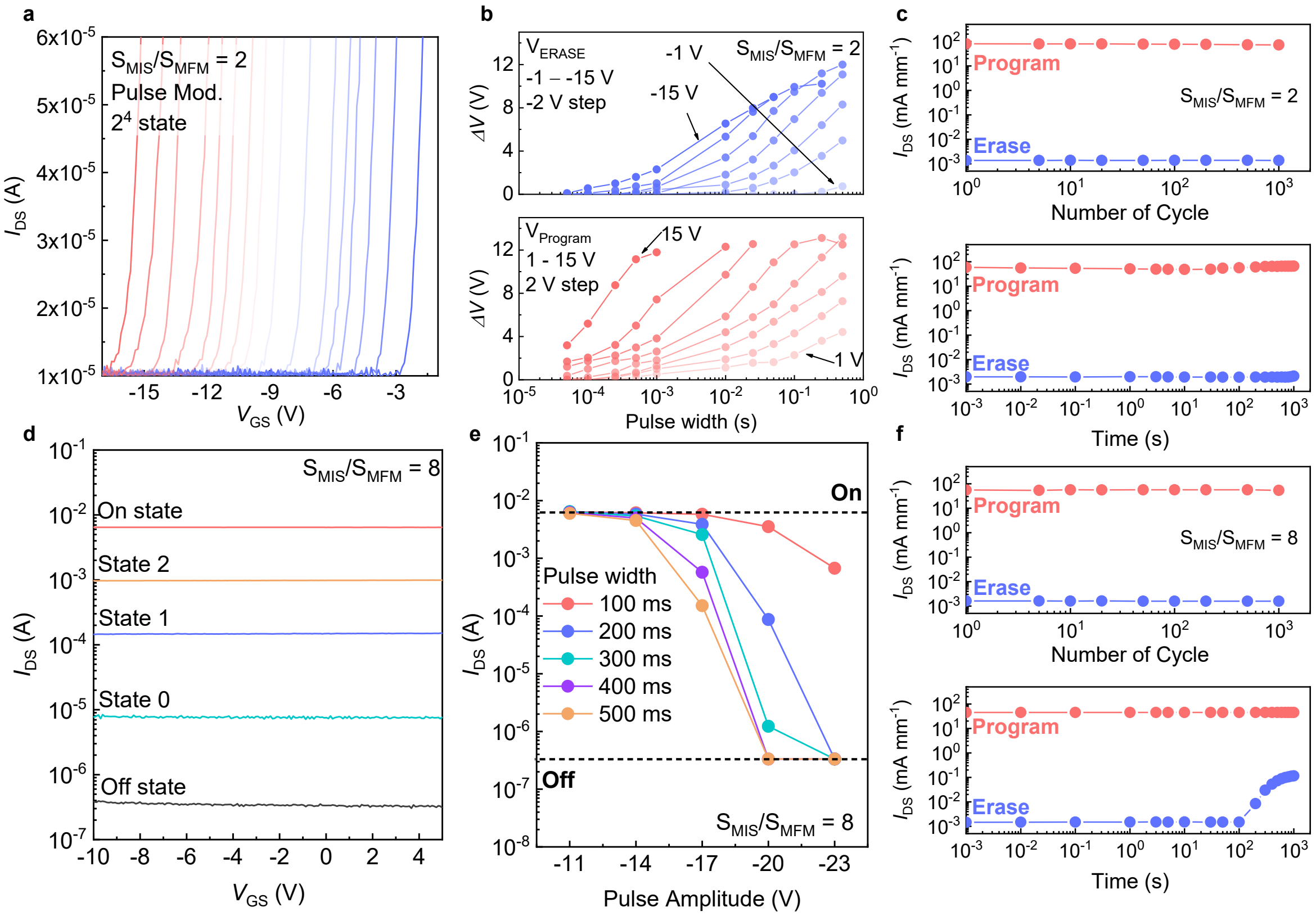

**Figure 3. Multi-bit storage capabilities and reliability of FeHEMTs. (a)** Drain current-gate voltage curves showing 16 discrete states (4-bit) achieved through gate pulse modulation for area-ratio 2. **(b)** Extracted ΔV results under various gate pulse amplitudes and widths for area-ratio 2. **(c)** Robust endurance up to 1,000 cycles and stable retention characteristics exceeding 1,000 s for area-ratio 2. **(d)** Drain current-gate voltage curves showing 5 discrete states achieved through gate pulse modulation for area-ratio 8. **(e)** Extracted drain current results under various gate pulse amplitudes and widths for area-ratio 8. **(f)** Endurance for 1,000 cycles and retention degradation after 100 s for area-ratio 8.

Figure 3a presents pulse modulated conductance for the area-ratio of 2. The 4-bit multi-level cell (MLC) operation comprising 16 deterministically distinguishable states is clearly resolved. Dependence of MW on pulse amplitude and width is systematically mapped in Figure 3b (see Figures. S3a and S3b for the ratio of 1). For the ratio 2 device, it achieves a remarkably increased maximum MW of 12 V. Notably, a robust MW exceeding 6 V is attained even under a lower amplitude pulse of 3 V, and full polarization saturation is effectively realized within a microsecond regime. In contrast, the ratio 1 device exhibits slower domain dynamics, as sub-millisecond pulse widths are insufficient to trigger significant polarization switching. Furthermore, its maximum MW saturates at approximately 6.3 V. This corroborates that area-ratio scaling accelerates the ferroelectric switching. Long-term reliability for the ratio of 2 is presented in Figure 3c (see Figure S3c for the ratio of 1). These results demonstrate that both endurance and non-volatile retention are highly robust. Specifically, these devices maintain stable and distinguishable states without severe polarization fatigue up to 1,000 switching cycles, while securely retaining the programmed data for over 1,000 seconds.

For the high area-ratio of 8, memory characteristics are evaluated through discrete 2DEG conductance levels rather than conventional full-swing transfer curves as shown in Figure 3d (see Figure S3d for the ratio of 4). Under transient pulses, these devices exhibit continuous, analog-like conductance modulation. This behavior originates from the robust, pre-induced downward polarization in the AlScN layer. This establishes a strong electrostatic barrier that prevents 2DEG depletion during pulse read operations. Consequently, these high-ratio devices operate as highly linear, multi-state conductance modulators. Each discrete state corresponds to a precisely controlled fraction of partial polarization reversal. Figure 3e present erase pulse modulation for the ratio of 8 (see Figure S3e for the ratio of 4). Program pulse responses are not shown separately because the heavily pre-aligned downward polarization yields a single saturated state rather than progressive modulation. The analysis reveals that the ratio 8 device necessitates a higher erase pulse amplitude compared to ratio 4. This indicates that the higher $Q_{2DEG}$ in the ratio 8 device exerts a stronger electrostatic screening effect, thereby pinning the downward polarization more rigidly and increasing the effective coercive field for the erase operation. Finally, the long-term reliability is presented in Figure 3f and Figure S3f. These devices demonstrate robust endurance up to 1,000 cycles. However, they suffer from noticeable retention degradation of the Erase state after 100 seconds. The observed retention degradation is due to the high $Q_{2DEG}$ generating a persistent built-in depolarization field that restores the upward-aligned domains to their original stable, pre-pinned downward state. It is amplified at the high area-ratios because the $Q_{2DEG}$ to be counterbalanced increases with $L_{Pt}$. Moreover, conduction into the floating Pt electrode can contribute to neutralizing the field that the polarization imposes on the 2DEG channel, further weakening the Erase state. This trade-off defines a functional divergence; the low-ratio devices (1 and 2) ensure robust non-volatile retention suited to high-density MLC storage, while high-ratio devices (4 and 8) provide unique merits for precise, continuously tunable multi-state conductance for analog switching and signal processing.

Figure 4a presents bidirectional transfer curves obtained under variable $V_{DS}$, for the area-ratio 2 FeHEMT (see Figure S4a for the ratio 1). A well-defined counterclockwise hysteresis loop and a consistently maintained MW are observed at $V_{DS}$ of as low as 0.5 mV. Figure 4b and Figure S4b quantify the extracted key device parameters, confirming their invariance across the measured $V_{DS}$ range. The preservation of robust ferroelectric switching even at a record-low $V_{DS}$ of 0.5 mV demonstrates the exceptional potential of this architecture for low-power and energy-constrained applications.

To verify scalability, 4×4 memory arrays were fabricated and evaluated (Figure 4c and Figure S4c). The gate width/length (W/L) of individual cell was 12 μm/7 μm. It is to be noted that the area-ratios of 1 and 2 were selected for the array demonstration on the basis of their superior retention stability relative to the high ratio devices. In Figure 4d, the bidirectional transfer curves for all 16 individual cells of the ratio 2 array are overlaid. Figure 4e illustrates the related pulse modulation characteristics, clearly showing a wide and robust read margin of 6 V across the pulse responses. Figure 4f provides a detailed view of the quantitative statistical distributions and the spatial MW color chart, which highlights the excellent uniformity between cells. Similarly, the memory characteristics of the ratio 1 array are thoroughly assessed. Figure S4d displays the highly consistent DC transfer curves. Additionally, Figure S4e shows the superimposed pulse modulation responses, which confirm a read margin of 4.2 V. The integrated statistical analysis and the corresponding MW color chart for this setup are shown in Figure S4f. These results collectively demonstrate the viability of large-scale GaN-based NVM.

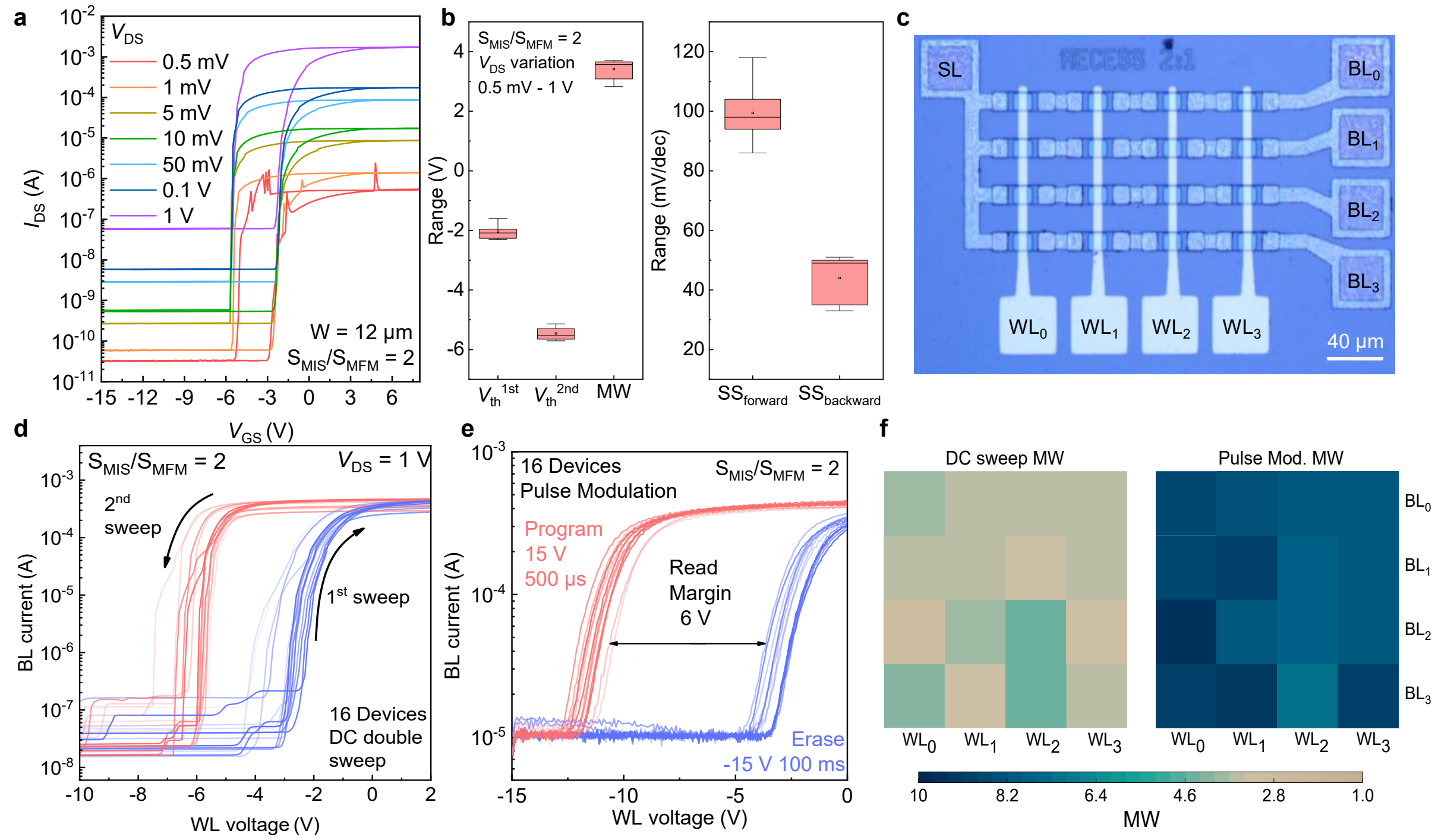

**Figure 4. Uniformity analysis and spatial mapping of 4×4 FeHEMT array with area-ratio 2.** **(a)** Bidirectional transfer characteristics of the FeHEMT with area-ratio 2 under varying $V_{DS}$ conditions. **(b)** Statistical box plots of the extracted 1st $V_{th}$, 2nd $V_{th}$, MW, $SS_{forward}$, and $SS_{backward}$ for area-ratio 2. **(c)** Optical microscope image of the fabricated 4×4 array with area-ratio 2. **(d)** Bidirectional transfer curves for the 4×4 array with area-ratio 2 (16 individual devices). **(e)** Gate pulse modulation responses for the 4×4 array with area-ratio 2. **(f)** Color-coded MW spatial heatmaps (DC sweep and pulse modulation) for the 4×4 array with area-ratio 2.

We performed a comprehensive benchmarking analysis in comparison with established channel materials for FeNAND, such as Si and IGZO (Table S1). This rigorously demonstrates competitiveness of the proposed FeHEMT. Notably, the device with an area-ratio of 2 exhibits a maximum MW of 12 V, which not only rivals but also surpasses the performance metrics of traditional Si- and IGZO-based technologies. This study illustrates the capability to maintain a broad MW under an exceptionally low $V_{DS}$ of 5 mV. This is accompanied by energy-efficient operation due to reduced channel current levels. The pulse amplitudes and widths necessary for reliable programming are comparable to those of silicon and oxide semiconductors. These findings position the FeHEMT as a promising high-performance alternative to silicon and oxide semiconductors for next-generation and harsh environment non-volatile memory applications.

Departing from conventional binary logic inverters, we demonstrate the first multi-state GaN inverter, which can improve information-processing density while reducing circuit complexity and footprint (Figures. S5a and S5b). To evaluate the capability for multi-state inverter, voltage transfer curves (VTCs) of the high area-ratio devices (4 and 8) were measured at $V_{DD}$ = 5 V with a 1 MΩ load resistor (Figures. S5c and S5d). The VTCs exhibit a distinct clockwise hysteretic behavior, which originates from the counterclockwise hysteresis of the FeHEMT. This ferroelectricity-induced hysteretic behavior enables the continuous tuning of the channel resistance, which directly translates into programmable multi-state inverter outputs. Schematic diagrams of the multi-state inverter and its analog circuit application of a frequency-to-voltage converter (FVC) are illustrated in Figure S6. The proposed system leverages the gradual polarization switching of the AlScN layer to generate adjustable output voltages, which can be controlled by modifying the input pulse parameters. This feature enables the inverter to function as a crucial component in analog signal processing, where the output voltage directly reflects the input frequency. These multi-level output states are attributed to the capability of high area-ratio devices to exhibit precisely quantized channel conductance states under transient pulse regimes. By precisely adjusting the pulse duration applied to the gate of driver transistor, the channel resistance can be tuned across a wide dynamic range. Figures 5a–5f validate this progressive modulation: As the "Set" pulse duration increases from 50 ms to 400 ms, the AlScN polarization is driven further toward the upward-aligned state. This electrostatically depletes the 2DEG charges, resulting in a monotonic increase in the driver's channel resistance and thus $V_{OUT}$ from 0 V to 4.7 V accordingly.

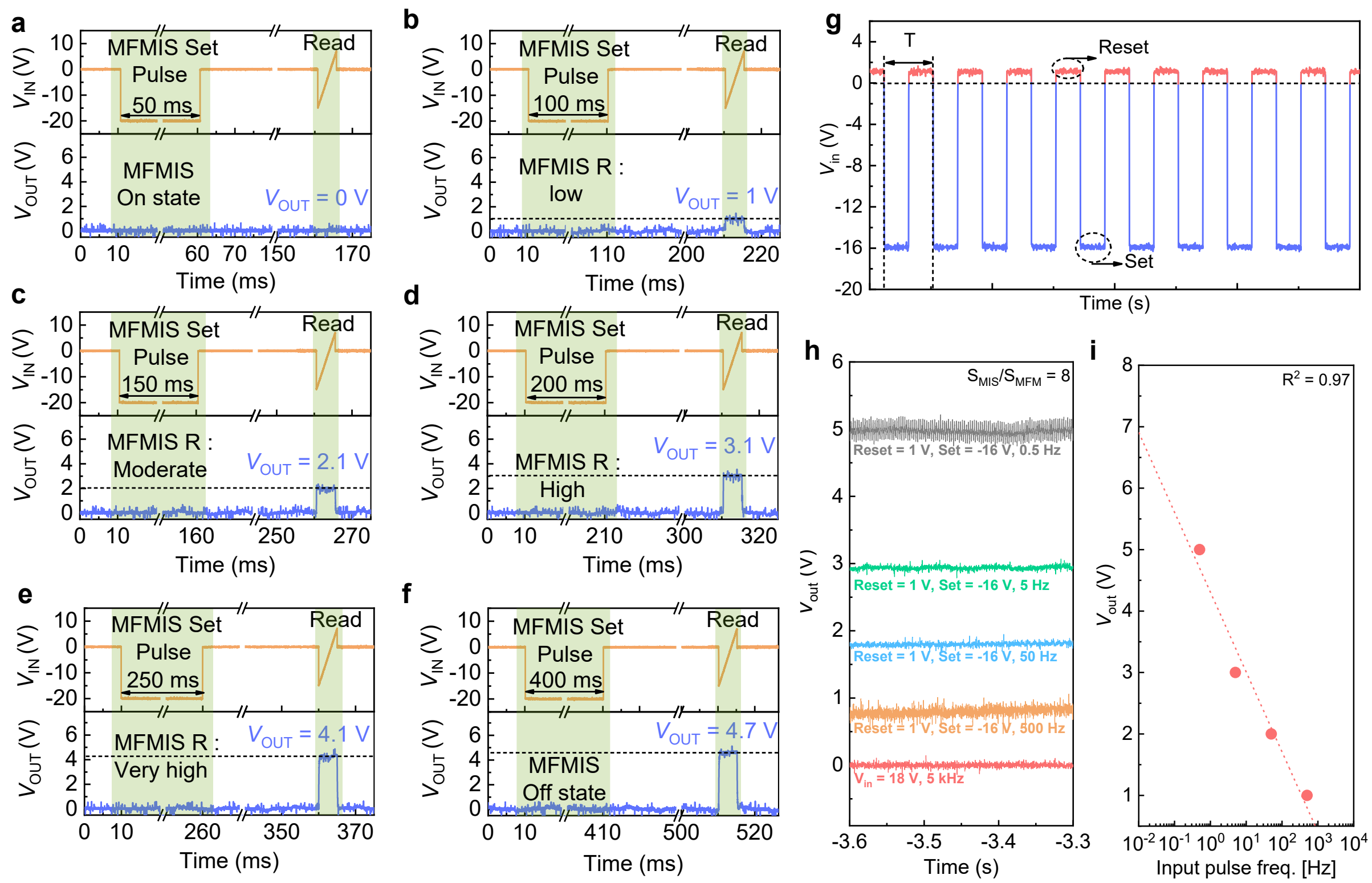

**Figure 5. Operational characteristics of a multi-state inverter and high-precision frequency-to-voltage converter based on FeHEMT. (a-f)** Progressive modulation of the multi-state inverter, exhibiting discrete output voltage levels ($V_{OUT}$) from 0 V to 4.7 V achieved by adjusting the "Set" pulse duration (50 ms to 400 ms). **(g)** Operating principle of the FVC utilizing a bipolar pulse scheme (1 V for Reset and -16 V for Set) to ensure operational stability. **(h)** Measured transient output waveforms demonstrating stable signal conversion across input pulse frequencies from 0.5 to 500 Hz. **(i)** Linear F-V transfer curve on a semi-logarithmic scale confirming the high precision of the FVC ($R^2$ = 0.97).

As a proof-of-concept for FeHEMT-based analog circuits, we demonstrate the first GaN-based FVC employing the proposed multi-state inverter. The FVC is a fundamental analog circuit block widely used in sensor interfaces, phase-locked loops, and motor control systems to translate frequency-encoded signals into proportional voltage outputs[34]. Conventional FVCs rely on charge-pump or switched-capacitor architectures that require multiple active components; in contrast, the proposed approach encodes frequency information directly into fractional ferroelectric polarization states of a single transistor, substantially reducing circuit complexity. As illustrated in Figure S6, the FVC converts the frequency of an input signal into a proportional output voltage. To evaluate this functionality, input pulse trains with frequencies ranging from 0.5 to 500 Hz were applied. As depicted in Figure 5g, a bipolar pulse scheme comprising both "Reset" and "Set" pulses is strictly required to ensure operational stability. The optimized amplitudes for two phases are Set as 1 V and -16 V, respectively. In the absence of a Reset phase, the continuous accumulation of Set pulses would drive the AlScN polarization into complete upward saturation. Therefore, a periodic Reset operation is imperative to re-initialize the polarization and maintain the dynamic equilibrium corresponding to the target frequency state. Figure 5h displays the measured transient output waveforms depending on variable input frequencies with same Reset/Set amplitudes, and four different $V_{OUT}$ were obtained from four input frequencies. Representative full transient input pulses and corresponding $V_{OUT}$ of 2 V are shown in Figure S7. Different pulse amplitudes are required to achieve the same $V_{OUT}$ depending on the area-ratios. This observation demonstrates highly stable signal conversion, featuring a calculated conversion gain of 1.1 mV/Hz at 500 Hz. Notably, the high linearity ($R^2 = 0.97$) of the frequency-to-voltage transfer curve on a semi-logarithmic scale, as plotted in Figure 5i, rigorously validates the precision of the analog output encoded by fractional polarization states. By establishing the area-ratio engineering as a structural design principle rather than a material dependent constraint, the ferroelectric GaN transistors enable a single platform route by effectively bridging the border between high-density non-volatile memory and analog signal processing systems.

In summary, we have demonstrated that systematic area-ratio ($S_{MIS}/S_{MFM}$) engineering of the AlScN-based MFMIS gate stack provides a unified design strategy for ferroelectric GaN transistors, simultaneously enabling record non-volatile memory performance and analog signal-processing functionality on a single platform. The coupling between the 2DEG and AlScN polarization plays the key role: High $Q_{2DEG}$ pins a robust downward ferroelectric polarization whose reversal drives a sub-Boltzmann subthreshold swing during the forward sweep. Increasing the area-ratio redistributes the gate bias voltage almost across the AlScN layer, achieving a record MW of 27 V and accelerating pulsed switching to the microsecond regime. In the memory regime, low area-ratio devices achieve 4-bit multi-level cell operation at a record-low $V_{DS}$ of 0.5 mV with excellent 4×4 array uniformity, benchmarking favorably against silicon and IGZO ferroelectric NAND architectures. In the analog regime, high area-ratio devices exploit continuously tunable conductance states to realize the first GaN-based ferroelectric frequency-to-voltage converter with exceptional linearity ($R^2$ = 0.97) over 0.5–500 Hz. Although retention in the high-ratio devices requires further gate-stack optimization, this work establishes area-ratio engineering as a versatile and practical design principle for wide-bandgap ferroelectric electronics, opening a route towards monolithically integrated GaN systems that unify high-density memory and precision analog processing.

**Supporting Information**

Experiment sections (fabrication of ferroelectric MFMIS GaN HEMTs, material and device characterization); comparison of DC and memory characteristics between silicon and oxide-based FeNAND devices and the proposed ferroelectric GaN transistor; schematic of the top view, optical microscope image of fabricated FeHEMT, and MFM capacitor characteristics; area-ratio dependent transfer characteristics; memory characteristics and reliability of the FeHEMTs with an area-ratio 1 and 4; uniformity and mapping of 4×4 FeHEMT array with an area-ratio of 1; schematic illustration of conventional and multi-state inverters, and measured voltage transfer curves of the multi-state inverters; schematic of the multi-state inverter using the MFMIS FeHEMT and its application to a frequency to voltage converter; transient pulse waveforms of frequency-to-voltage converter.

**Author Contributions**

HJ, GY conceived the idea. HJ, GY designed the experiment. HJ, JP worked on device fabrication and characterization. HJ, JP, KS, BH, GY conducted analysis. HJ, BH, GY wrote the first draft of the manuscript with input from all the authors. HJ, BH, GY reviewed and revised the manuscript. All the authors discussed the results and approved the manuscript.

**Notes**

The authors declare that they have no competing interests.

**Acknowledgements**

This work was supported by the National Research Foundation of Korea (NRF) grant (RS-2024-00438811, RS-2024-00455423) and Korea Evaluation Institute of Industrial Technology (KEIT) grant (RS-2024-00509266).

Supporting Information

# Monolithic GaN Systems Combining Non-Volatile Memory and Analog Computing via Area-Ratio-Engineered Ferroelectric AlScN Gate Stacks

*Hyeong Jun Joo[1], Jong Min Park[1], Kieran Barrett-Snyder[2], Brendan Hanrahan[3,*], and Geonwook Yoo[1, 4, *]*

**Corresponding authors. Email: brendan.m.hanrahan.civ@army.mil, gwyoo@ssu.ac.kr*

[1] Department of Intelligent Semiconductor, Soongsil University, Seoul 06938, Korea

[2] Fibertek Inc, Herndon, VA, USA

[3] DEVCOM Army Research Laboratory, Adelphi, MD, USA

[4] School of Electronic Engineering, Soongsil University, Seoul 06938, Korea

**Table of contents**

Figure S5. Schematic illustration of conventional and multi-state inverters, and measured voltage transfer curves of the multi-state inverters.

Figure S6. Schematic of the multi-state inverter using the MFMIS FeHEMT and its application to a frequency to voltage converter.

Figure S7. Transient pulse waveforms of frequency-to-voltage converter.

**Experimental sections**

**Fabrication of ferroelectric MFMIS GaN HEMTs** The AlGaN/GaN epitaxial layers utilized in this work were grown on (111) Si substrates through metal-organic chemical vapor deposition (MOCVD). The epitaxial stack is composed of a 250 nm AlN nucleation layer, a 2.5 μm GaN buffer, a 150 nm GaN channel, and a 1 nm AlN spacer. A 25 nm $Al_{0.26}Ga_{0.74}N$ barrier and a 3 nm GaN capping layer were sequentially integrated to complete the heterostructure. Device processing commenced with the realization of the source and drain ohmic contacts. A Ti/Al/Ni/Au (25/140/40/50 nm) multi-layer metal stack was patterned using electron-beam evaporation, followed by rapid thermal annealing (RTA) at 870 °C for 30 seconds in an $N_2$ ambient. A gate recess process was then executed via inductively coupled plasma (ICP) etching with a $BCl_3/Cl_2$ gas mixture. For the gate dielectric, a 9 nm $HfO_2$ insulating layer was deposited by atomic layer deposition (ALD) using TEMAH and $O_3$ as the hafnium precursor and reactant, respectively. To form the Metal-Insulator-Semiconductor (MIS) configuration, a 50 nm Pt intermediate metal layer was evaporated. Subsequently, a 50 nm $Al_{0.7}Sc_{0.3}N$ ferroelectric thin film was synthesized by reactive RF-magnetron sputtering using an $Al_{0.57}Sc_{0.43}$ alloy target. The sputtering conditions included an RF power of 280 W, a pressure of 5 mTorr, and a substrate temperature of 400 $^{o}$C under an $Ar/N_2$ environment. To conclude the fabrication, contact vias for the source and drain were defined using ICP etching, and a 50 nm Ni electrode was deposited to form the Metal-Ferroelectric-Metal (MFM) gate stack. The physical dimensions of the characterized devices were a channel length of 15 μm, and a gate length of 7 μm.

**Material and device characterization** Cross-sectional structure of the MFMIS gate stack was examined using high-resolution transmission electron microscopy with JEM-2100F(JEOL). DC electrical characteristics were measured using a Keithley 4200A-SCS(Keithley) parameter analyzer equipped with preamplifiers. Pulsed measurements were performed using a 4225-PMU(Keithley) with an amplifier unit. The output voltage waveforms of the multi-state inverter and FVC were captured using an MSO44, 4-BW-350(Tektronix) oscilloscope.

**Table S1.** Comparison of DC and memory characteristics between FeNAND devices and the proposed ferroelectric GaN transistor

| | VLSI'24 [29] | IEDM'25 [30] | IEDM'24 [31] | Sci.Adv.'21 [32] | Nature'25 [33] | This Work $S_{MIS}/S_{MFM} = 2$ |
|---|---|---|---|---|---|---|
| Channel | IGZO | IGZO | Si | $InZnO_x$ | IGZO | GaN |
| $V_{DS}$ [V] | 0.1 | - | 1 | 0.01 | - | 0.005 |
| $I_{on}$ [A] | $> 10^{-7}$ | $> 10^{-5}$ | $> 10^{-3}$ | $> 10^{-7}$ | $> 10^{-6}$ | $10^{-5}$ |
| $I_{off}$ [A] | $< 10^{-10}$ | $< 10^{-6}$ | $< 10^{-6}$ | $< 10^{-10}$ | $< 10^{-9}$ | $3 \times 10^{-10}$ |
| Program Voltage [V] | 7-11 | 8-18 | 11-15 | 5 | 5-15 | 1-15 |
| Pulse width [μs] | 1-100 | 1-1,000 | 1-500 | 10,000 | 1-10,000 | 50-50,000 |
| Max MW [V] | 17.8 | 9.4 | 12.2 | 2.5 | > 10 | 12 |

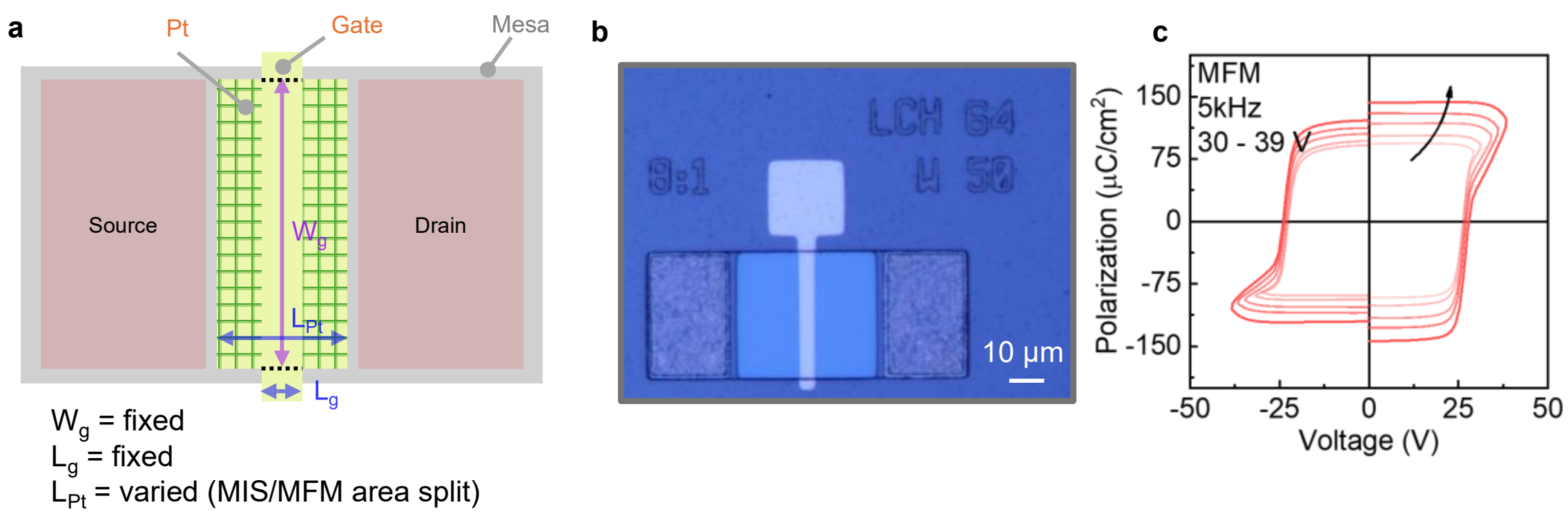

**Figure S1. Schematic of the top view, optical microscope image of fabricated FeHEMT, and MFM capacitor characteristics. (a)** Top-view schematic of the fabricated device. **(b)** OM image of the MFMIS FeHEMT with an area-ratio of 8. **(c)** P-V hysteresis curves extracted using the Positive Up Negative Down (PUND) method, demonstrating the progressive transition toward full polarization saturation under increasing applied fields.

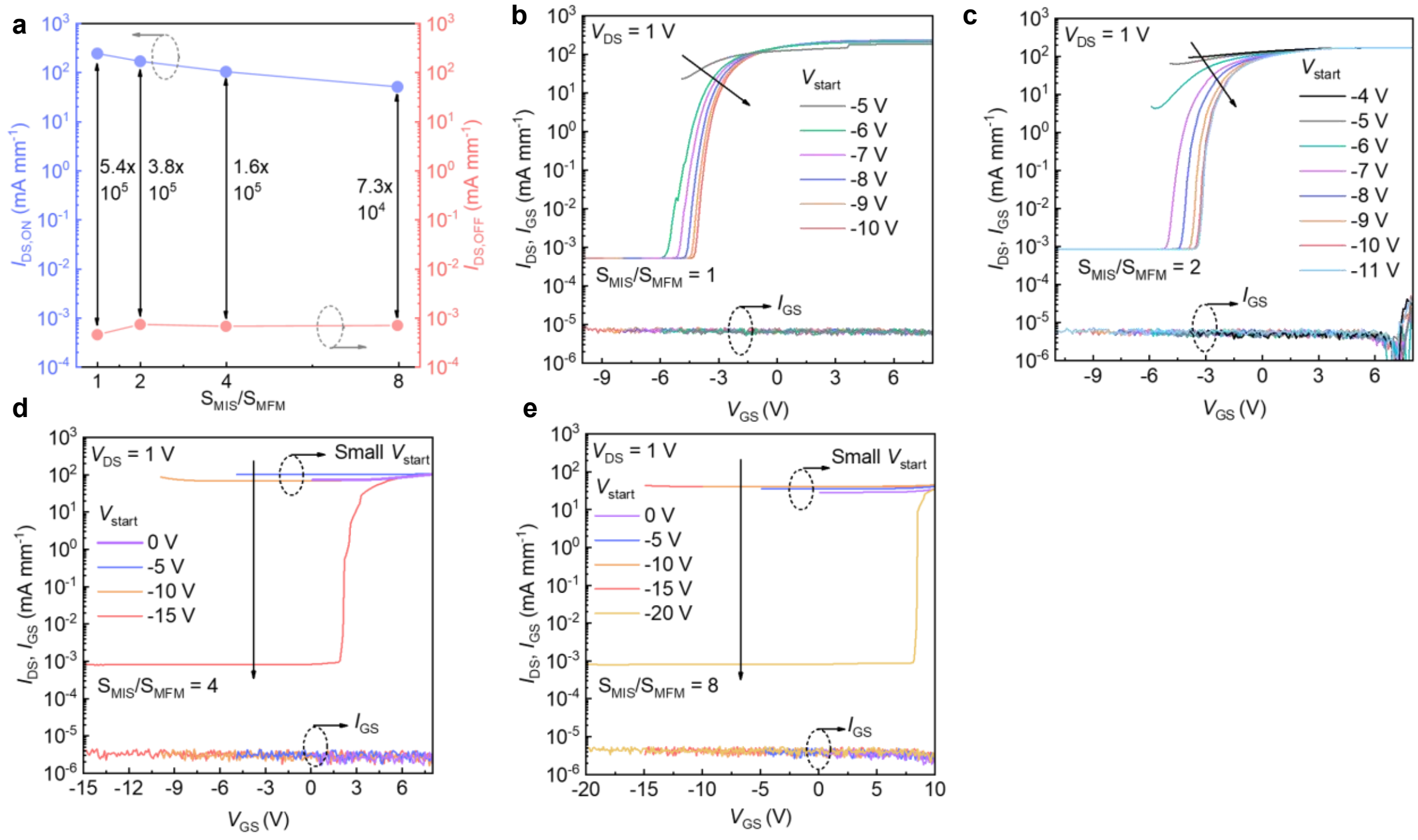

**Figure S2. (a)** Extracted $I_{on}/I_{off}$ ratio as a function of area-ratio. **(b-e)** Transfer curves and device behavior under varying starting gate voltages $V_{start}$ for area-ratios 1 (b), 2 (c), 4 (d), and 8 (e).

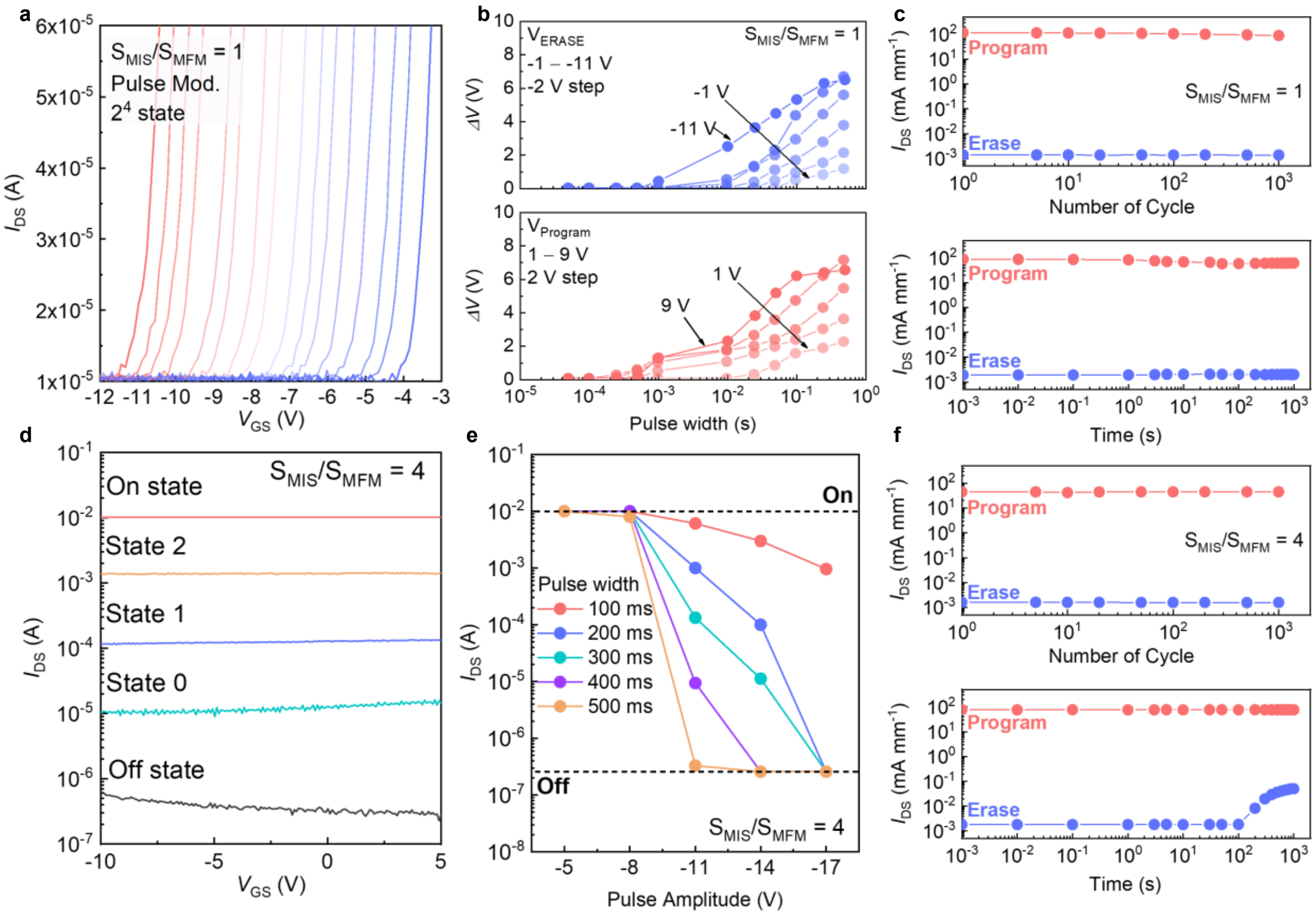

**Figure S3. Multi-bit storage capabilities and reliability of fabricated FeHEMTs. (a)** Drain current ($I_{DS}$) vs. gate bias ($V_{GS}$) curves showing 16 discrete states (4-bit) achieved through gate pulse modulation for area-ratio 1. **(b)** Extracted ΔV results under various gate pulse amplitudes and widths for area-ratio 1. (**c**) Stable retention characteristics exceeding 1,000 s and robust endurance up to 1,000 cycles for area-ratio 1. **(d)** Drain current-gate voltage curves showing 5 discrete states achieved through gate pulse modulation for area-ratio 4. **(e)** Extracted drain current results under various gate pulse amplitudes and widths for area-ratio 4. **(f)** Endurance for 1,000 cycles and retention degradation after 100 s for area-ratio 4.

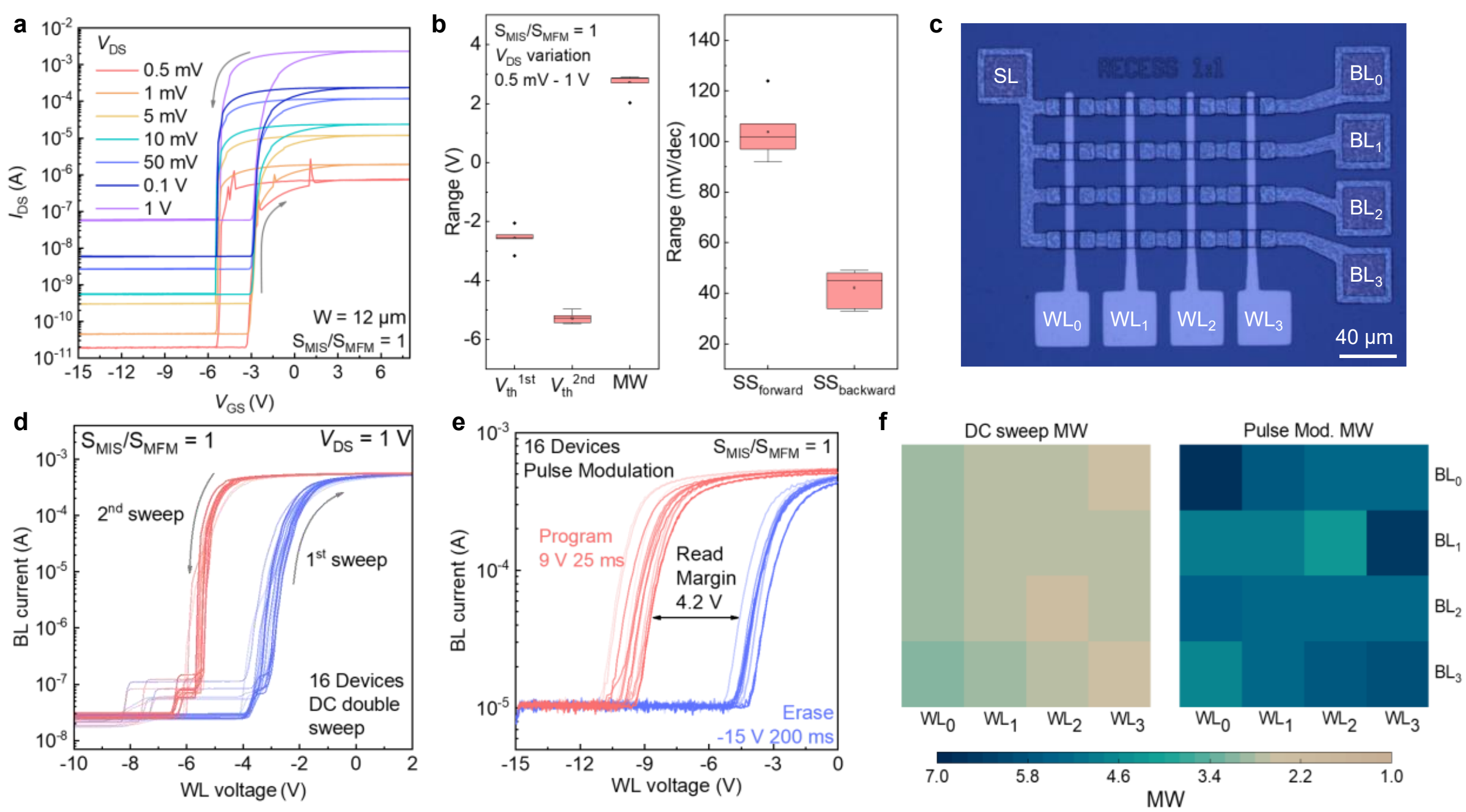

**Figure S4. Uniformity analysis and spatial mapping of 4×4 FeHEMT array. (a)** Bidirectional transfer characteristics of the FeHEMT with area-ratio 1 under varying $V_{DS}$ conditions. **(b)** Statistical box plots of the extracted 1st $V_{th}$, 2nd $V_{th}$, MW, $SS_{forward}$, and $SS_{backward}$ for area-ratio 1. **(c)** Optical microscope image of the fabricated 4×4 array with area-ratio 1. **(d)** Bidirectional transfer curves for the 4×4 array with area-ratio 1 (16 individual devices). **(e)** Gate pulse modulation responses for the 4×4 array with area-ratio 1. **(f)** Color-coded MW spatial heatmaps (DC sweep and pulse modulation.) for the 4×4 array with area-ratio 1.

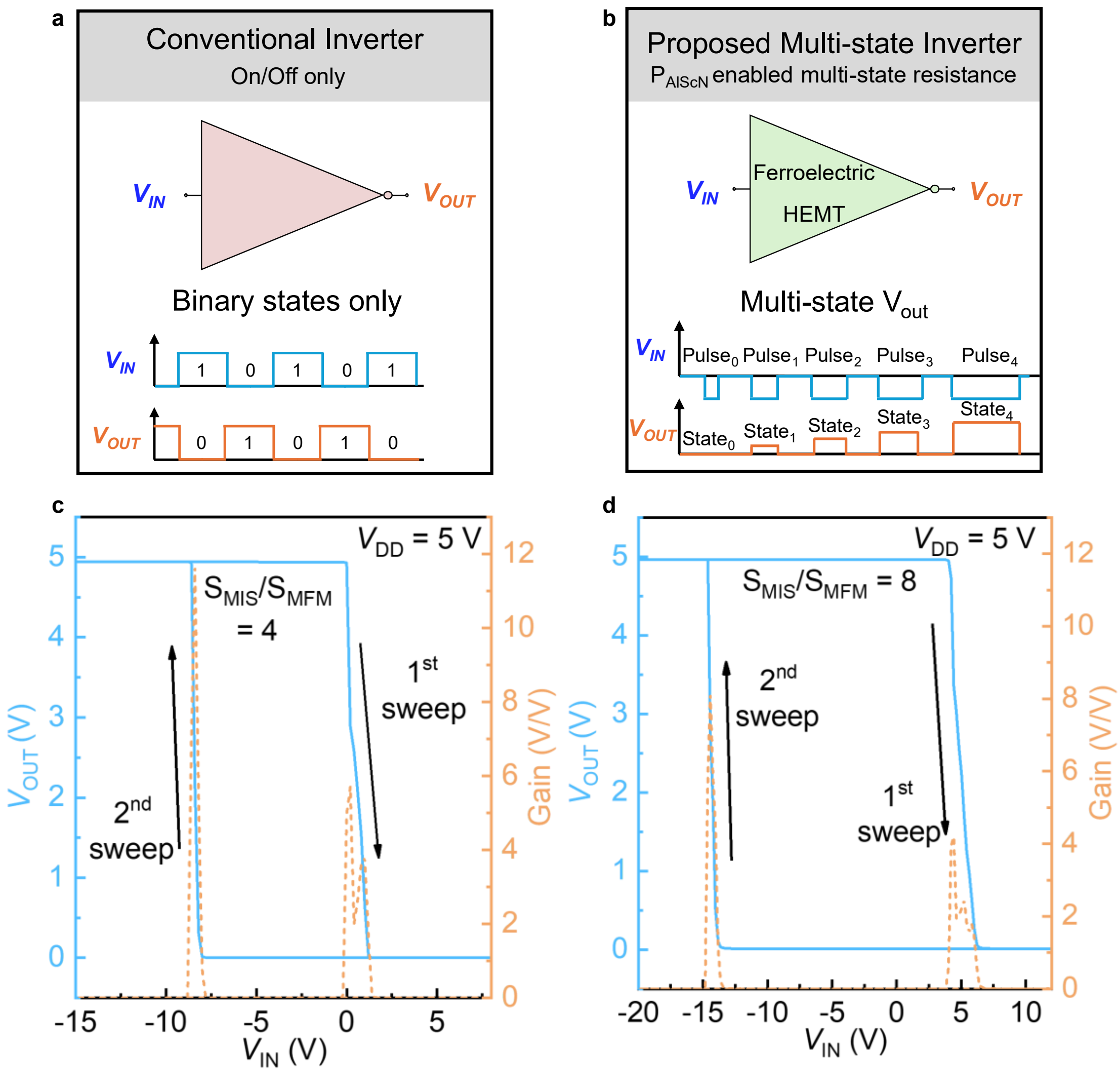

**Figure S5. Schematic illustration and transfer voltage curves of the multi-state inverters.** Operational comparison between **(a)** a conventional logic inverter and **(b)** the proposed reconfigurable multi-state FeHEMT inverter. **(c), (d)** Measured DC voltage transfer characteristics and pronounced hysteresis for area-ratios of 4 and 8.

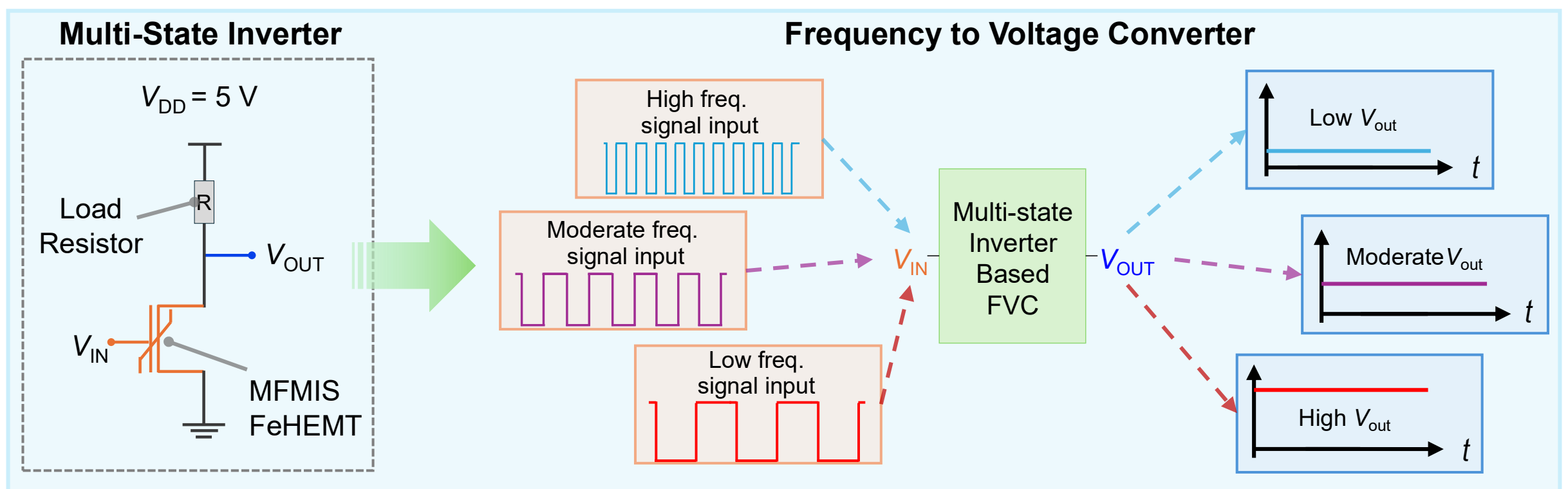

**Figure S6.** Schematic of the multi-state inverter using the MFMIS FeHEMT and its application to a frequency to voltage converter.

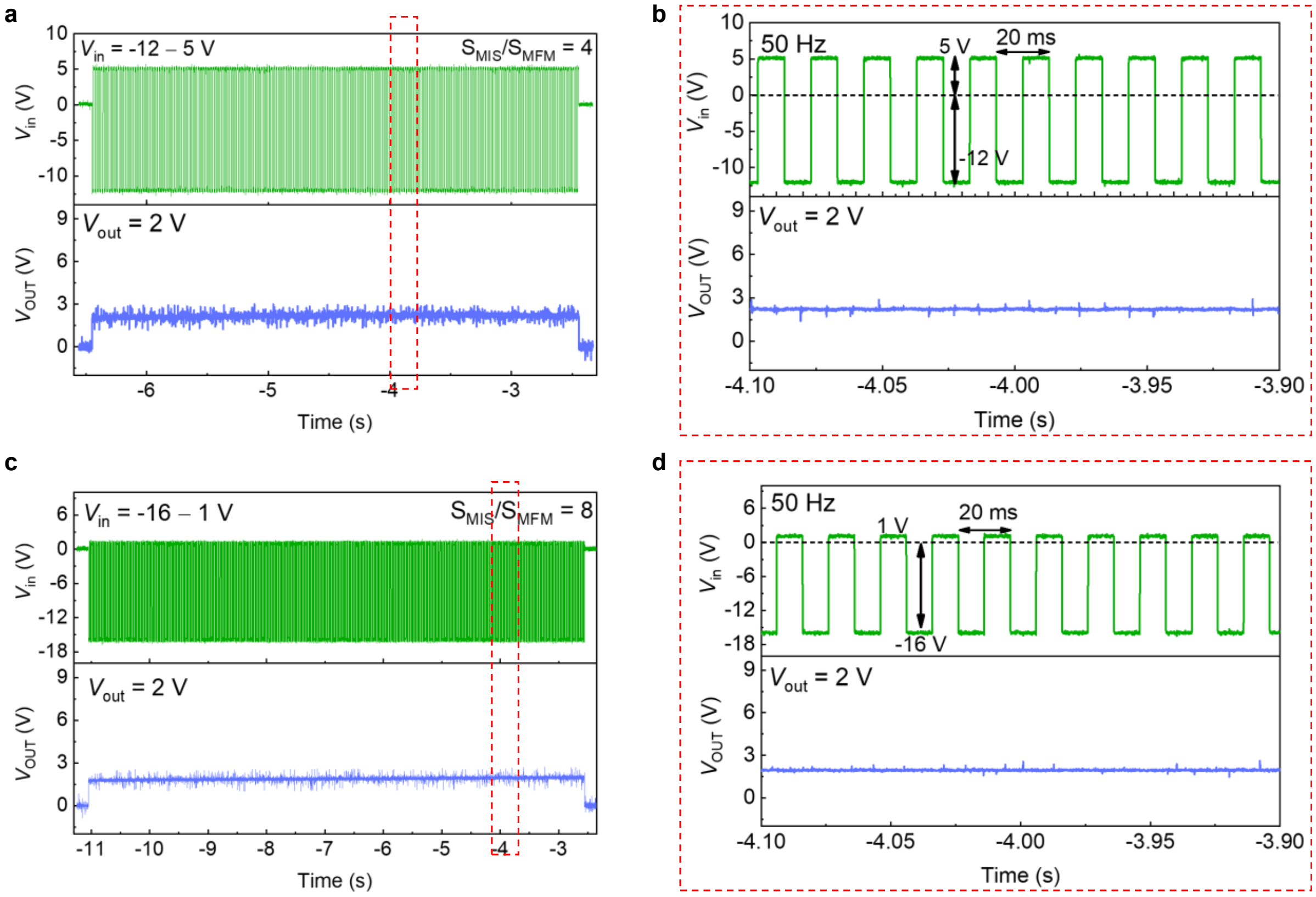

**Figure S7. Transient pulse waveforms for target voltage generation.** Complete input pulse sequences and the resulting $V_{OUT}$ for achieving a target level of 2 V. **(a), (b)** Full waveform and its magnified view for the area-ratio 4 device. **(c), (d)** Full waveform and its magnified view for the area-ratio 8 device.